\documentclass[%
 reprint,
 floatfix,
 amsmath,amssymb,
 aps,
]{revtex4-2}
\usepackage{float}
\usepackage{graphicx}
\usepackage{dcolumn}
\usepackage{bm}
\usepackage{xcolor}

\begin{document}

\preprint{APS/123-QED}

\title{Study of Baryon Number Transport Dynamics and Strangeness Conservation Effects Using $\Omega$-hadron Correlations}

\author{Weijie Dong$^{1}$}
\author{Xiaozhou Yu$^{1}$}
\author{Siyuan Ping$^{1}$}

\author{Xiatong Wu$^{2}$}

\author{Gang Wang$^{2}$}
 \email{gwang@physics.ucla.edu}

\author{Huan Zhong Huang$^{1,2}$}
 \email{huang@physics.ucla.edu}

\author{Zi-Wei Lin$^{3}$}

\affiliation{$^{1}$Key Laboratory of Nuclear Physics and Ion-beam Application (MOE), Fudan University, Shanghai 200433, China  }%

\affiliation{$^{2}$
Department of Physics and Astronomy, University of California, Los Angeles, CA 90095, USA
}

\affiliation{$^{3}$
Department of Physics, East Carolina University, Greenville, NC 27858, USA
}

\date{\today}

\begin{abstract}

In nuclear collisions at RHIC energies, an excess of $\Omega$ hyperons over $\overline{\Omega}$ is observed, indicating that $\Omega$ carries a net baryon number despite $s$ and $\bar{s}$ quarks being produced in pairs. The baryon number in $\Omega$ could have been transported from the incident nuclei and/or produced in baryon-pair production of $\Omega$ with other types of anti-hyperons, such as $\overline{\Xi}$. To investigate these two scenarios, we propose to measure correlations between $\Omega$ and $K$, as well as between $\Omega$ and anti-hyperons. We will use two versions, the default and string-melting, of a multiphase transport (AMPT) model to illustrate the method to measure the correlation and to demonstrate the general shape of the correlation. We will present the $\Omega$-hadron correlations from simulated Au+Au collisions at $\sqrt{s_{NN}} =$ 7.7 and 14.6 GeV, and discuss the dependence on collision energy and on the hadronization scheme in these two AMPT versions. 
These correlations can be used to explore the mechanism of baryon number transport and the effects of baryon number and strangeness conservation in nuclear collisions. 

\end{abstract}

\maketitle
 
\vspace{.25in}

\section{INTRODUCTION}

Strangeness enhancement was proposed as a signature of the quark-gluon plasma (QGP) created in relativistic heavy-ion collisions~\cite{koch1986strangeness} and has been a subject of intensive theoretical and experimental investigations~\cite{adams2005experimental}. Since the incident protons and neutrons are composed of $u$ and $d$ quarks, the strange quarks observed in the aftermath of the collision can only originate from pair of $s\bar{s}$ production. Lattice Quantum ChromoDynamics (QCD) predicted that the temperature at which the quark-hadron phase transition occurs is approximately 150 MeV. Thus, in the QGP phase where the temperature is comparable to the $s$-quark mass, strangeness may be abundantly produced via flavor creation ($qq \rightarrow s\bar{s}$, $gg \rightarrow s\bar{s}$) and gluon splitting ($g \rightarrow s\bar{s}$), leading to enhanced production of strangeness in the final state~\cite{koch1986strangeness,rafelski1982strangeness}. The investigation of multi-strange hyperon production is particularly valuable for studying the equilibration of strangeness in the QGP. The yields of $\Omega$ hyperons in heavy-ion collisions, for example, have been measured to be significantly higher than those in $p$+$p$ collisions scaled by the number of participants at 
CERN Super Proton Synchrotron (SPS)~\cite{WA97:1999ikd}, Relativistic Heavy Ion Collider (RHIC)~\cite{STAR:2005gfr,STAR:2007cqw}, and the Large Hadron Collider (LHC)~\cite{ALICE:2013xmt,ALICE:2016fzo}. Furthermore, many strange hadrons are believed to have small hadronic rescattering cross-sections, so that they  retain information from the hadronization stage and can be used to probe the phase boundary of the quark-hadron transition~\cite{van1998evidence,takeuchi2015effects,mohanty2009probe}.

\begin{figure}[bthp]
\centering
\graphicspath{ {figures/} }
\includegraphics[width=0.3\textwidth]{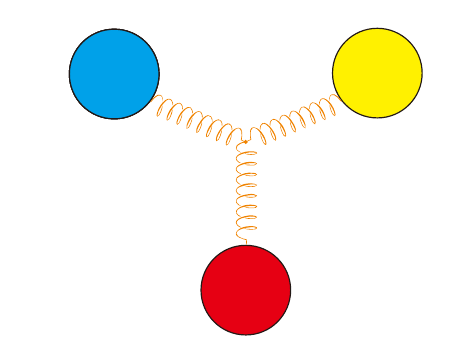}
\caption{\label{fig:wide} Gluon junction configuration is the Y-shaped gluon field (orange curve) in the structure of the baryon which is speculated to be the baryon number carrier. The three circles represent valance quarks in a baryon.
\label{fig:0}
}
\end{figure}

\begin{figure*}[bthp]
\centering
\graphicspath{ {figures/} }
\includegraphics[width=0.9\textwidth]{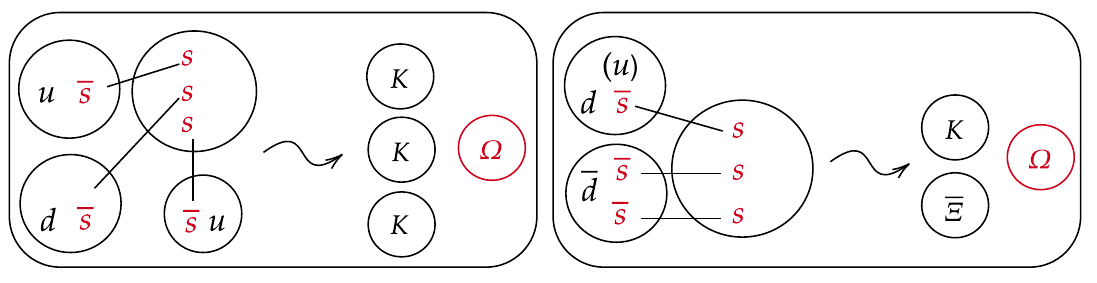}
\caption{\label{fig:wide2} Schematic illustration of two possible scenarios for the $\Omega^{-}$ production based on the quark coalescence picture with strangeness and baryon number conservation. In Scenario 1 (left),  $\Omega^{-}$ carries the baryon number initially residing in the $u$ and $d$ quarks from colliding nuclei. In Scenario 2 (right),  $\Omega^{-}$ does not carry a net baryon number in the production mechanism. Associated production of $\Omega$ and kaon in AMPT default version could have similar qualitative features as Scenario 1 in coalescence picture. The $\bar{\Xi}$ hyperon in the figure is an example of pair production of $\Omega$ and $\bar{\Xi}$.
\label{fig:1}
}
\end{figure*}

The RHIC Beam Energy Scan (BES) program aims to search for a possible critical point in the QCD phase diagram where strangeness production also plays a major role~\cite{STAR:2019bjj}. At  low BES energies, the measured ratios of anti-baryons to baryons are significantly lower than unity at midrapidities for three hyperons ($\Lambda$, $\Xi$, and $\Omega$). Therefore, these hyperons must carry a net baryon number. The baryon number transport dynamics have been a subject of interest  of the heavy-ion collision physics since its inception~\cite{busza1984nuclear,buszaLedoux1988energy,huang1999selected}. The baryon number in a proton or neutron may be attributed to  valence $u$ and $d$ quarks, each carrying $1/3$ of the baryon number. However, in high energy collisions the valence quarks tend to inherit a significant fraction of the incident nucleon momentum,  making them ineffective in transporting baryon number from beam rapidities to midrapidities. Previous theoretical calculations have proposed the interactions of topological objects known as gluon junctions shown in FIG. 1, which can effectively convey baryon number over large rapidity gaps in nuclear collisions~\cite{kharzeev1996can}. Once a gluon junction reaches  midrapidities during a nuclear collision, it must emerge as a baryon in the final state, with its flavor  determined by the quark flavors present in the surrounding QGP medium.
Such exotic dynamics have a particularly discernible impact on net $\Omega$ hyperons, as $\Omega$ hyperons consist of three $s$ quarks that must be pair-produced.  

The quantum numbers of strangeness and baryon number are strictly conserved in nuclear collisions, leading to correlations among particles in the final state. To gain insights into the production dynamics of $\Omega$ hyperons, we propose to measure the correlations between $\Omega$ and other particles, namely $K^+$, $\bar{\Lambda}$, and $\bar{\Xi}^{+}$, along with their respective anti-particle pairs. By examining the shape and strength of these correlation functions, we illustrate the extent that we can quantitatively characterize the role of conservation laws in the $\Omega$ production dynamics in nuclear collisions. These correlations will also provide a means to search for possible exotic dynamics such as gluon junction interactions for baryon number transport.

In this paper, we use a multiphase transport (AMPT) model~\cite{lin2005multiphase} to simulate Au+Au collisions at $\sqrt{s_{NN}} =$ 7.7 GeV and 14.6 GeV and present the $\Omega$-hadron correlations from these simulations. In section 2, we describe briefly the AMPT simulations and the analysis methods. The correlation results and discussions are presented in section 3, followed by a summary in section 4.

\section{AMPT Simulations and Analysis Methods}

\subsection{AMPT Simulations}

To investigate the impact of different hadronization schemes on baryon number transport dynamics, we exploit both the default and string melting (SM) versions of the AMPT model to simulate Au+Au collisions. The initial phase space is provided by the HIJING model~\cite{wang1991role,wang1991hijing,wang1992systematic,wang1992gluon}, which is a Monte Carlo event generator for parton and particle production in high-energy hadronic and nuclear collisions. In the default version of AMPT, minijets and their parent nucleons form excited strings after partonic interactions of minijets, and these strings fragment 
into hadrons via Lund String Fragmentation~\cite{andersson1983parton}. In the SM version, the strings are converted through the "string melting" mechanism into partons, which subsequently interact with each other during the evolution, and the coalescence mechanism is used to combine partons into hadrons. 
The parton interactions are described
by the Boltzmann equation, and solved by the ZPC model~\cite{li1995formation}, which only includes two-body elastic scatterings. In the quark coalescence process, the  two or three nearest partons (quarks and anti-quarks) in the phase space recombine to form a meson or a baryon. 
The AMPT version used in this paper, which strictly conserves net electric charge, strangeness, and baryon number, is employed in our study. We have generated more than 50 million minimum bias events of Au+Au collisions at both 7.7 GeV and 14.6 GeV.

 \subsection{$\Omega$ Production and Conservation of Strangeness and Baryon Number}

The $\Omega^-$ production is governed by both Strangeness Conservation (SC) and baryon number conservation. Additionally, the presence of net $\Omega$ hyperons at midrapidities indicates the involvement of Baryon Number Transport (BNT) dynamics.  Figure~\ref{fig:1} illustrates two basic scenarios based on the quark coalescence picture for the $\Omega^-$ production. In Scenario 1 (left side of Fig.~\ref{fig:1}), 
three $\bar s$ quarks are pair-produced with the three $s$ quarks in
$\Omega^{-}$, and may combine with $u$ or $d$ quarks from the incident 
nuclei to form three kaons. In this case, the $\Omega^-$ hyperon does not have an accompanying anti-baryon, and it carries a fraction of the total net baryon number, which is transported from the colliding nuclei. In exotic dynamics involving gluon junction interactions, the production of the three $s$-$\bar{s}$ pairs may be combined with a gluon junction, which gives rise to $\Omega^-$. Consequently, Scenario 1 is inclusive of possible exotic BNT via gluon junction to the $\Omega$ hyperon, while valence $u/d$ quarks form the kaons~\cite{huang1999selected}. In default version, associated production of $\Omega$ and kaons could also yield qualitatively similar features as the coalescence picture. Therefore, Scenario 1 encompasses contributions from both ordinary physical processes and  the gluon junction dynamics. The AMPT model in our simulation does not explicitly include the gluon junction dynamics. We use Scenario 1 to denote general process where a net baryon number is transported to the $\Omega$ hyperon regardless the underlying dynamics. 
In Scenario 2 (right side of Fig.~\ref{fig:1}), the three pair-produced $\bar{s}$ quarks  combine with $u$($\bar{u}$) or $d$($\bar{d}$) quarks to form an anti-hyperon and a kaon. For instance,  $\bar{\Xi}^{+}(\bar{\Xi}^{0})$ and $K^{0}(K^{+})$ could emerge alongside the $\Omega^{-}$ hyperon. In this scenario, the baryon number in  $\Omega^-$ is balanced by the other types of  anti-hyperon, and no net baryon number from the incident nuclei is present. 

To characterize the numbers of kaons and anti-baryons associated with the $\Omega$ production, we introduce  
\begin{equation}
\label{eq:1}
\Delta N_A \equiv \langle A \rangle_{{\rm w.}\Omega^{-}} - \langle A \rangle_{{\rm w.o.}\Omega^{-}},
\end{equation}
where $\langle A \rangle_{{\rm w.}\Omega^{-}}$ and $\langle A \rangle_{{\rm w.o.}\Omega^{-}}$ denote the average numbers of particle $A$ in events with one $\Omega^-$ and without any $\Omega^-$, respectively.
Here, we assume that all other aspects of the two event classes are the same. Table~\ref{table:syst_glw1} lists the $\Delta N_{K}$ and $\Delta N_{\bar{B}}$ values expected by the two scenarios, where $K$ represents both $K^{+}$ and $K^{0}$, and $\bar{B}$ refers to anti-baryons ($\bar{\Lambda}$, $\bar{\Sigma}$, $\bar{\Xi}$, and so on) associated with the $\Omega^{-}$ production. In our study, we choose the beam energies of 7.7  and 14.6 GeV in order to limit the fraction of Au+Au events that produce multiple $\Omega$ ($\bar\Omega$) hyperons. Compared with Scenario 2, Scenario 1 exhibits a stronger $\Omega^{-}$-$K$ correlation and a weaker $\Omega^{-}$-$\bar{B}$ correlation.

 \begin{table}[htb]
 \caption{$\Delta N_K$ and $\Delta N_{\bar B}$ in the two scenarios for the production of an $\Omega^{-}$ .}
 \label{table:syst_glw1}

  \begin{tabular}{lccc}
  \hline
  \hline 
 &  $\Delta N_{K}$     & $\Delta N_{\bar B}$ \\ \hline
  Scenario 1 (SC+BNT)  & 3                   & 0   \\ \hline
  Scenario 2 (SC)      & 1, 2, 3               & 1 \\ 
  \hline
  \end{tabular}

 \end{table}

\begin{figure*}[htbp]
\graphicspath{ {figures/} }
\includegraphics[width=0.90\textwidth]{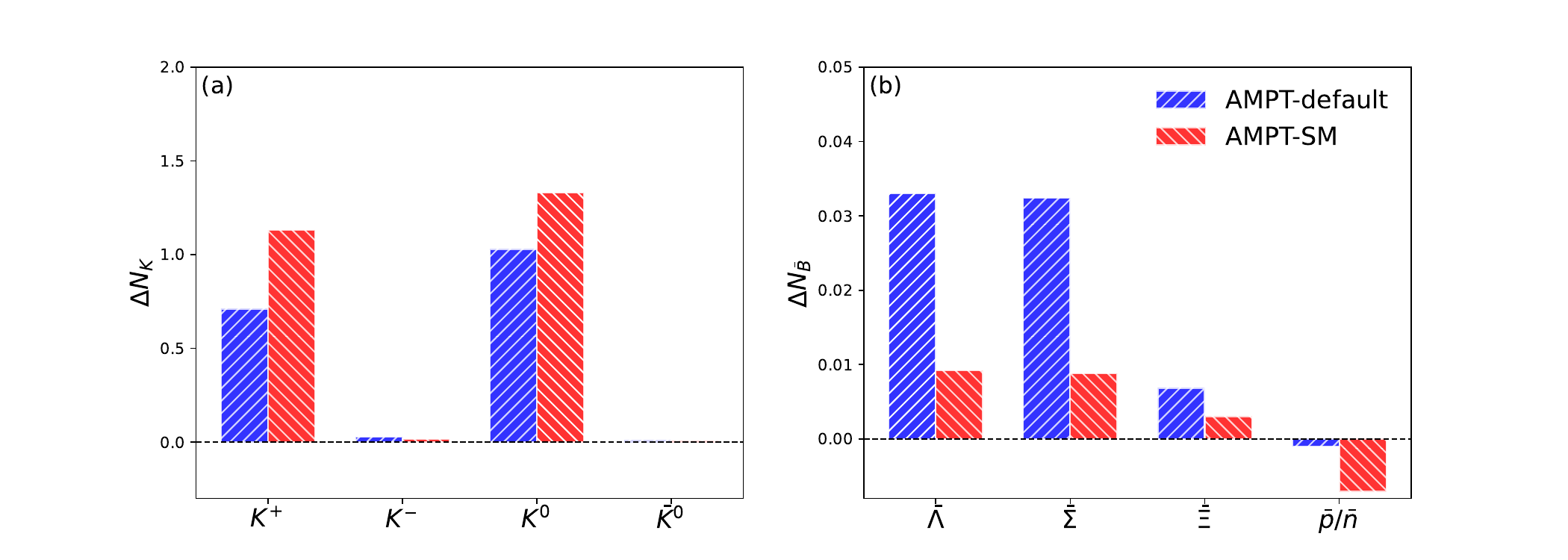}
\caption{\label{fig:2} 
AMPT calculations of the difference in strange hadron yields between $\Omega^{-}$ events and non-$\Omega^{-}$ events of Au+Au collisions at 7.7 GeV.
}
\end{figure*}

\begin{figure*}[htbp]
  \centering

\includegraphics[width=0.90\textwidth]{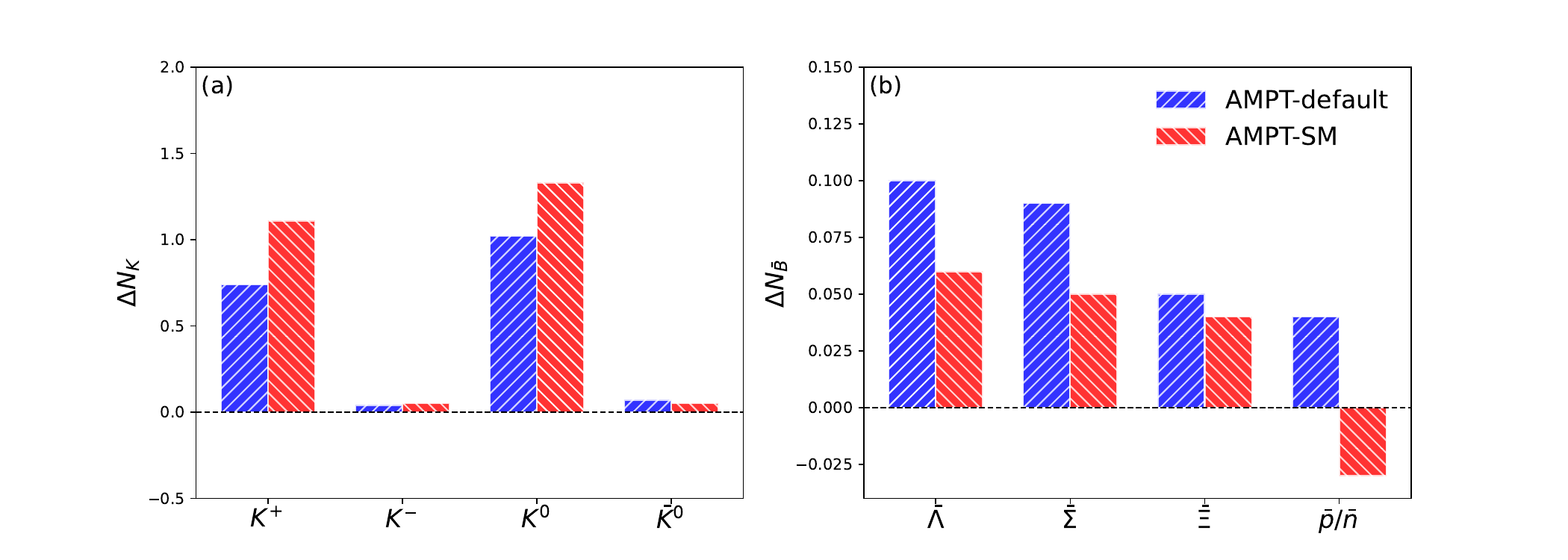}
\caption{\label{fig:3} 
AMPT calculations of the difference in strange hadron yields between $\Omega^{-}$ events and non-$\Omega^{-}$ events of Au+Au collisions at 14.6 GeV.
}
\end{figure*}

\begin{figure*}[htbp]
\graphicspath{ {figures/} }
\includegraphics[width=0.90\textwidth]{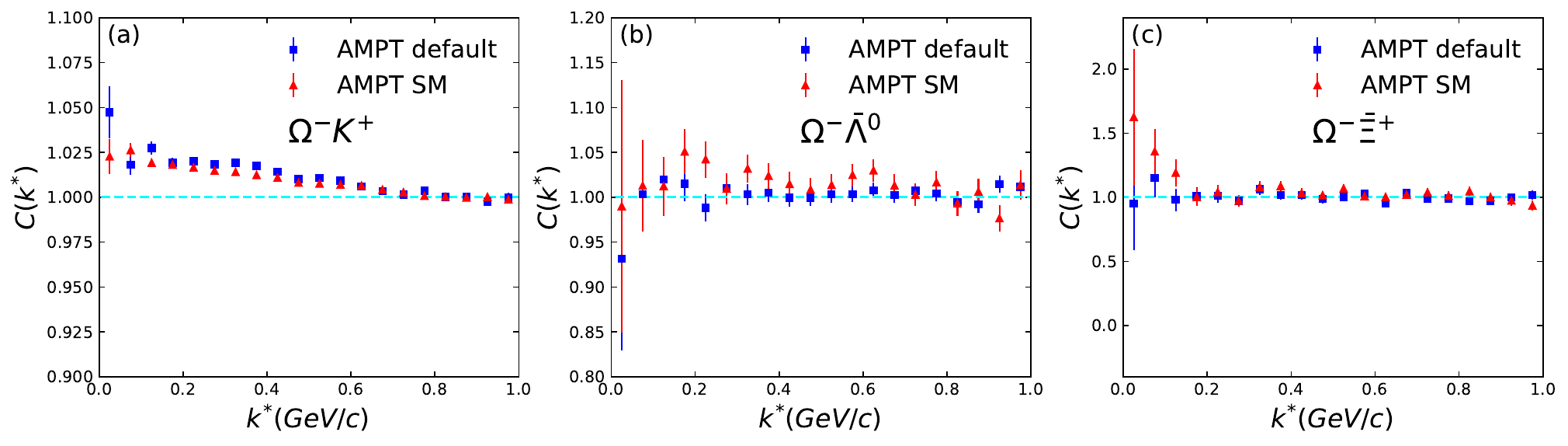}
\caption{\label{fig:4} Correlation functions between $\Omega^{-}$ and strange hadrons using the event mixing normalization in the 0--5\% centrality range of Au + Au collisions at 7.7 GeV ($|\eta|<1$).}
\end{figure*}

\begin{figure*}[htbp]
\graphicspath{ {figures/} }
\includegraphics[width=0.90\textwidth]{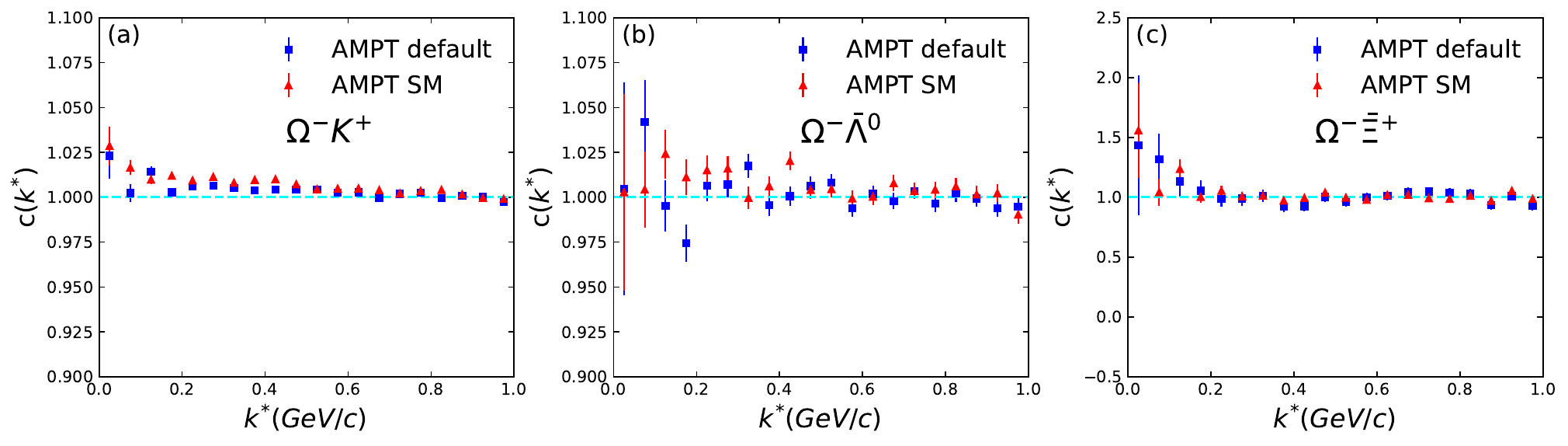}
\caption{\label{fig:5} Correlation functions between $\Omega^{-}$ and strange hadrons using the event mixing normalization in the 0--5\% centrality range of Au + Au collisions at 14.6 GeV ($|\eta|<1$).}
\end{figure*}

\subsection{Correlation Using the Event Mixing Technique}

To divide out the combinatorial background,
we first analyze the correlations between $\Omega^{-}$ and strange hadrons  with the traditional normalization using mixed events similar to the scheme used for other analyses, e.g., ~\cite{STAR:2004yym}. The correlation function is 
\begin{equation}\label{2}
    C(k^*)=\mathcal{N} \frac{A(k^*)}{B(k^*)},
\end{equation}
where $A(k^*)$ is the same-event distribution, $B(k^*)$ is the mixed-event distribution, and  $k^*\equiv\frac{1}{2}|k_1-k_2|$ is the reduced momentum in the pair rest frame. $\mathcal{N}$ is the normalization factor determined by matching the same-event and mixed-event correlations in the uncorrelated phase space, for example, by requiring $C(k^*>1$ GeV/$c)=1$. 

The technique of event mixing normalization allows for the investigation of the correlation length between two types of particles in momentum space by analyzing their distributions. The underlying assumption is that the distribution in mixed events should be equivalent to that in the same events in the absence of any correlation. By dividing out this mixed-event background, one can obtain the correlation between the two particles of interest. It is crucial to choose a proper $k^{*}$ range for normalization in order to achieve meaningful results. Typically, the $k^{*}$ range with the high counting density is selected for normalization, as it provides a reliable estimate of the background. This method has also been applied to examine the correlation function for $\Omega^{-}$-$K^{+}$, $\Omega^{-}$-$\bar{\Lambda^0}$, and $\Omega^{-}$-$\bar{\Xi}^+$ in terms of relative rapidity and relative transverse momentum. 
However, it is important to note that due to the normalization procedure, this correlation function cannot fully capture the magnitude of multiple kaons being correlated with the $\Omega$ production, nor can it adequately represent the sensitivity of the correlation to the production dynamics.

\subsection{Correlations Using Combinatorial Background Subtraction}

In order to make the correlation measurement sensitive to the number of associated hadrons, we introduce the combinatorial background subtracted (CBS) correlations, by taking the difference between the $\Omega$-hadron correlation and the $\bar{\Omega}$-hadron correlation, each normalized by the corresponding number of $\Omega$ or $\bar{\Omega}$ hyperons. For example, the $\Omega^{-}$-$K^{+}$ correlation is defined as
\begin{equation}\label{3}
C^\text{CBS}_{\Omega^{-}K^{+}}(k^*)=\frac{dN_{\Omega^{-}K^{+}}/dk^{*}}{N_{\Omega^{-}}} - \frac{dN_{\bar{\Omega}^{+}K^{+}}/dk^{*}}{N_{\bar{\Omega}^{+}}}.
\end{equation}
This background subtraction approach is intended to extract the main component of the correlation due to SC in the $\Omega^-$-$K^+$ pairs.
The  opposite-sign pair ($\Omega^{-}$-$K^{+}$ or $\bar{\Omega}^{+}$-$K^{-}$) distribution in the first term contains the signal, while the same-sign pair distribution in the second term ($\bar{\Omega}^{+}$-$K^{+}$ or $\Omega^{-}$-$K^{-}$) models the uncorrelated background. 
This  subtraction scheme is  sensitive to the difference in the number of kaons between events with $\Omega^{-}$ and events with $\bar{\Omega}^{+}$, as well as to the phase space distribution of the extra kaons. There may be variations of kinematic phase spaces for these pairs in nuclear collisions, and such effects can be mitigated with this normalization scheme and with fine collision centrality bins.  Similar approaches have been applied to study the correlations between $\Omega$ and other hadrons.

\section{Correlation Results and Discussions}

\subsection{Strangeness Conservation and Strange Hadron Yields}

We first list in Table~\ref{table:syst_glw} the AMPT simulations of the difference in the number of $s\bar{s}$ pairs 
between events with one $\Omega^{-}$ and events without any $\Omega^{-}$ or $\bar{\Omega}^{+}$ in Au+Au collisions at $\sqrt{s_{NN}} =$ 7.7 GeV and 14.6 GeV. For events with one $\Omega$, at least three $s\bar{s}$ pairs are produced because of Strangeness Conservation (SC). The AMPT results are slightly greater than three, and the excess indicates  that the underlying strangeness production may affect the probability for the $\Omega$ formation. Compared with the SM version of AMPT, the  default version  yields a slightly larger  number of $s\bar{s}$ pairs, as well as a smaller $\Omega$ formation probability. This correlation  is presumably due to the difference in the formation dynamics and/or the $s\bar{s}$ phase space. The AMPT simulations suggest that a large number of strange quarks per event are needed in order to favorably form an $\Omega^{-}$ hyperon in collisions at lower energies.

\begin{table}[htbp]
\caption{Difference in the number of $s\bar{s}$ pairs in the 0–5\% centrality 
between events with one $\Omega^{-}$ and events without any $\Omega^{-}$ or $\bar{\Omega}^{+}$. }
\label{table:syst_glw}
  \begin{tabular}{lccccc}
 \hline 
  \hline 
& 7.7 GeV     & 14.6 GeV   \\ \hline
  AMPT SM             & 3.22       & 3.07        \\ 
  AMPT default        & 3.25       & 3.16           \\
  \hline 
  \end{tabular}

 \end{table}

\begin{table}[htbp]
\caption{Average numbers of $\Delta N_{K}$ and $\Delta N_{\bar{B}}$ in the 0–5\% centrality as defined in Eq.~(\ref{eq:1}).  }
 \label{table:syst_glw3}
  \begin{tabular}{lccccc}
 \hline 
  \hline 
  &  \multicolumn{2}{c} {7.7 GeV} &  &  \multicolumn{2}{c} {14.6 GeV}                          \\
\hline
 & $\Delta N_{K}$     & $\Delta N_{\bar{B}}$ &  & $\Delta N_{K}$     & $\Delta N_{\bar{B}}$ \\ \hline
  AMPT SM             & 2.46       & 0.017   &  & 2.44       & 0.119      \\ 
  AMPT default        & 1.74       & 0.078   &   &  1.76      & 0.28         \\
  \hline 
  \end{tabular}

 \end{table}

Table~\ref{table:syst_glw3} shows the AMPT calculations of $\Delta N_{K}$ and $\Delta N_{\bar{B}}$ as defined in Eq.~(\ref{eq:1}). When compared with the expectations in  Table~\ref{table:syst_glw1} corresponding to the two $\Omega$ production scenarios, these numbers  indicate that the $\Omega$ production is likely to receive contributions from both scenarios. The  SM version seems to favor the dominance of Scenario 1, with the $\Delta N_{K}$ values  close to three and the lower $\Delta N_{\bar{B}}$ values.

Strangeness and  baryon number can be represented by various particle types in the final states of  nuclear collisions. For example, $\bar{s}$ quarks can exist in $K^+$ and $K^0$. The actual distributions of strangeness and baryon numbers in the final-state particles could be sensitive to nuclear dynamics and may also depend on beam energy. 
Figures~\ref{fig:2} and \ref{fig:3} show the AMPT results of the difference in strange hadron yields between $\Omega^{-}$ events and non-$\Omega^{-}$ events of Au+Au collisions at  7.7 GeV and 14.6 GeV, respectively. At 7.7 GeV and 14.6 GeV, the correlations between $\Omega^-$ and kaons are stronger in the SM version, but the correlations between $\Omega^-$ and anti-hyperons are stronger in the default version. 
For $\Omega^-$ events, the number of kaons is much larger than the number of anti-hyperons, suggesting that the  correlations between $\Omega^-$ and kaons are much stronger than those between $\Omega^-$ and anti-hyperons. These numbers seem to support that scenario 1 contributes more significantly to $\Omega$ yields in both versions.


\subsection{Correlations between $\Omega^\pm$ and Strange Hadrons}

We first show the correlations between $\Omega$ and strange hadrons as a function of the reduced momentum $k^*$ in the pair rest frame using the event mixing normalization.
Figure~\ref{fig:4} presents the correlation functions for (a) $\Omega^{-}$-$K^+$, (b) $\Omega^{-}$-$\bar{\Lambda}$, and (c) $\Omega^-$-$\bar{\Xi}^+$ pairs for 0--5\% centrality Au + Au collisions at 7.7 GeV. The normalization between the same event and the mixed event distributions is determined by the $k^*$ region of 0.6--1.5 GeV/$c$. There is a distinct correlation between $\Omega^-$ and $K^+$, indicating that SC must play a major role in the $\Omega^-$ and $K^+$ yields, with the typical correlation length on the order of less than $k^*$ of 0.5 GeV/$c$. The $\Omega^-$ may also be correlated with the anti-hyperons of $\bar{\Lambda}$ and $\bar{\Xi}^+$, but the current simulated results do not have sufficient statistics to be definitive.

\begin{figure}[bt]
\graphicspath{ {figures/} }
\includegraphics[width=0.48\textwidth]{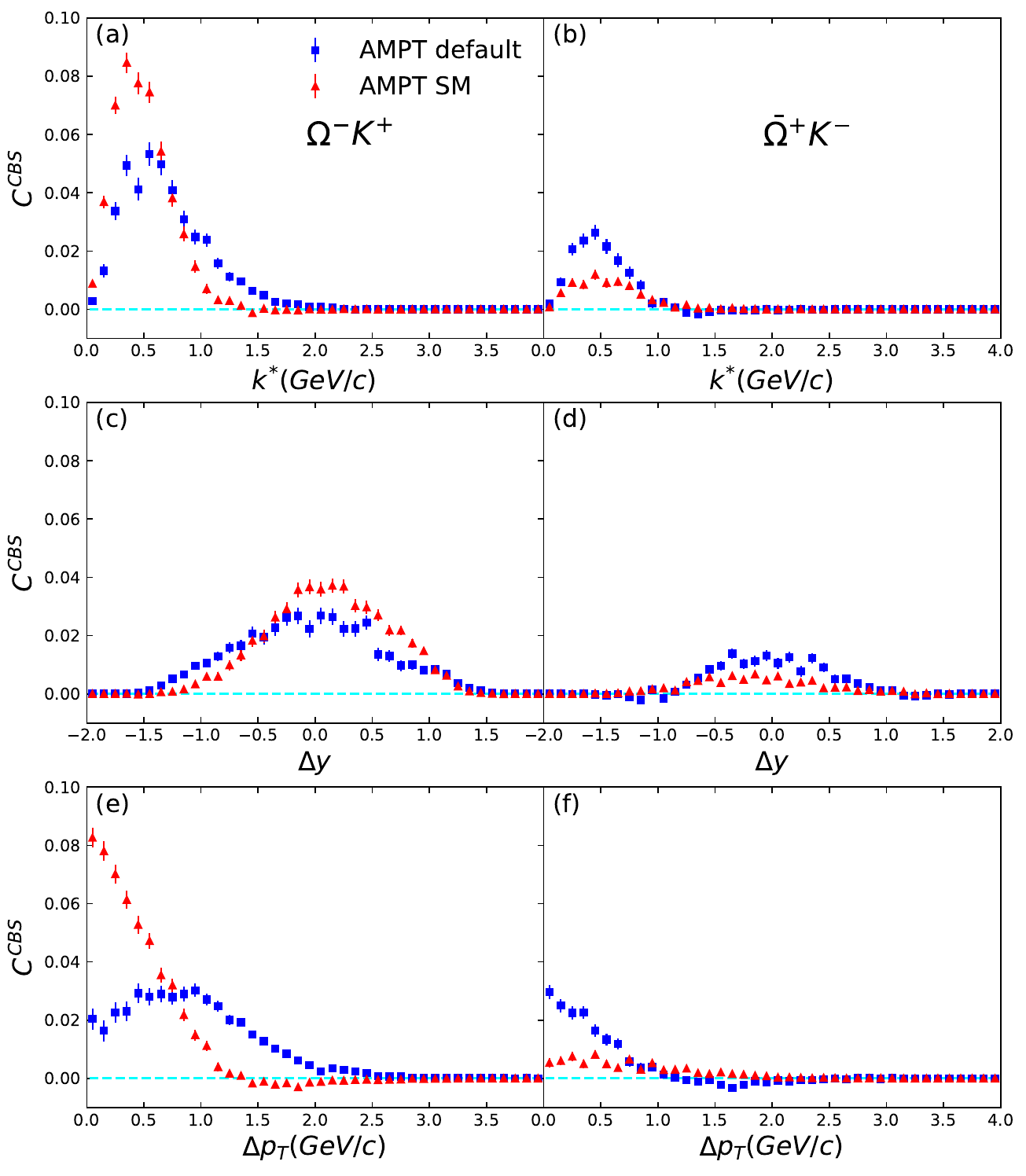}
\caption{\label{fig:6}
CBS correlations between $\Omega^-(\bar{\Omega}^{+})$ and kaons in 0--5\%
most central Au + Au collisions at 7.7 GeV ($|\eta|<1$). The left columns show the results of $C^{\rm CBS}_{\Omega^- K^{+}}$, and the right columns, those of $C^{\rm CBS}_{\bar{\Omega}^{+}K^{-}}$.
}
\end{figure}

\begin{figure}[bt]
\graphicspath{ {figures/} }
\includegraphics[width=0.48\textwidth]{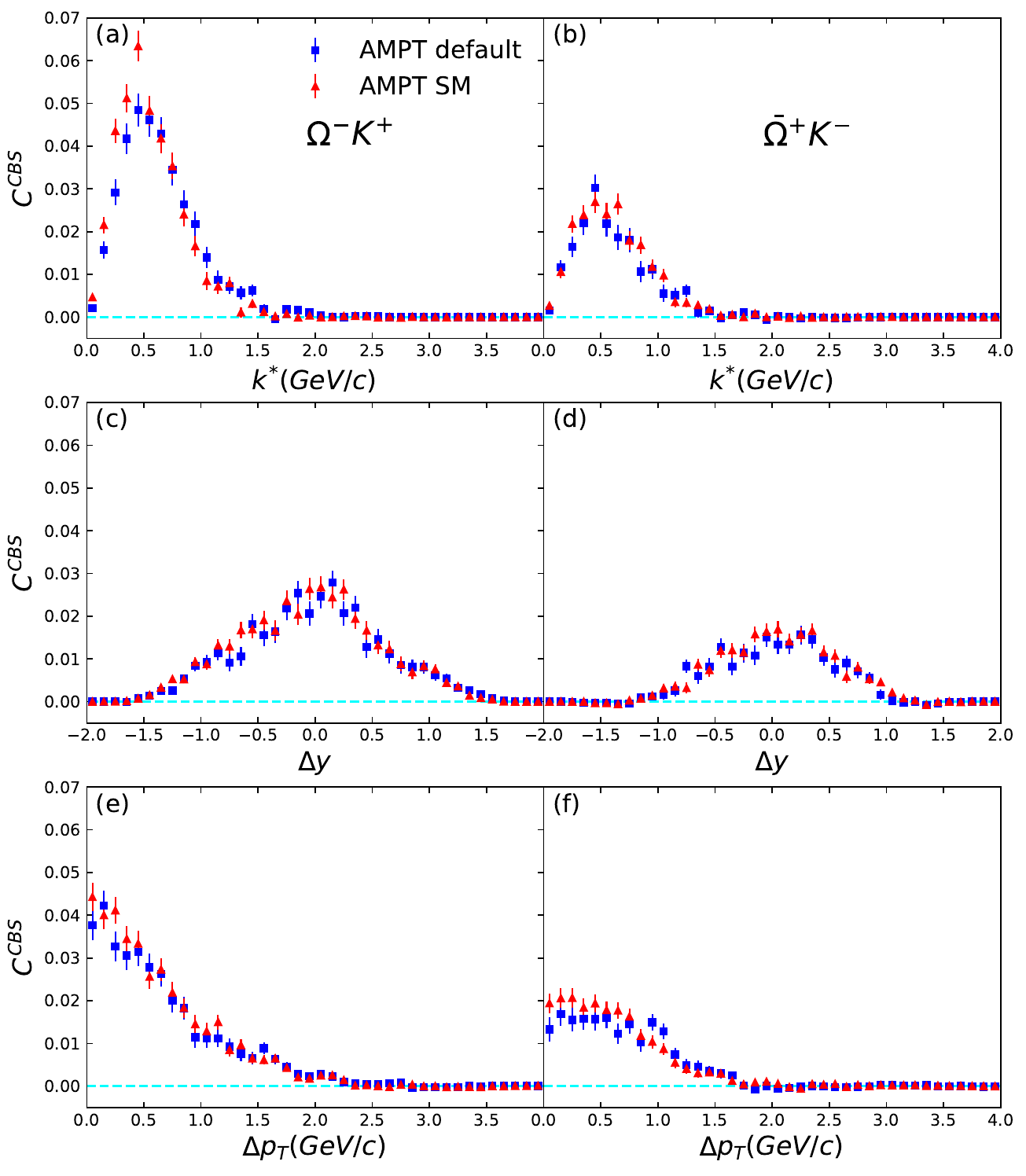}
\caption{\label{fig:7} CBS correlations between $\Omega^-(\bar{\Omega}^{+})$ and kaons in 0--5\%
most central Au + Au collisions at 14.6 GeV. The left columns show the results of $C^{\rm CBS}_{\Omega^- K^{+}}$, and the right columns, those of $C^{\rm CBS}_{\bar{\Omega}^{+}K^{-}}$.
}
\end{figure}

\begin{figure}[t]
\graphicspath{ {figures/} }
\includegraphics[width=0.48\textwidth]{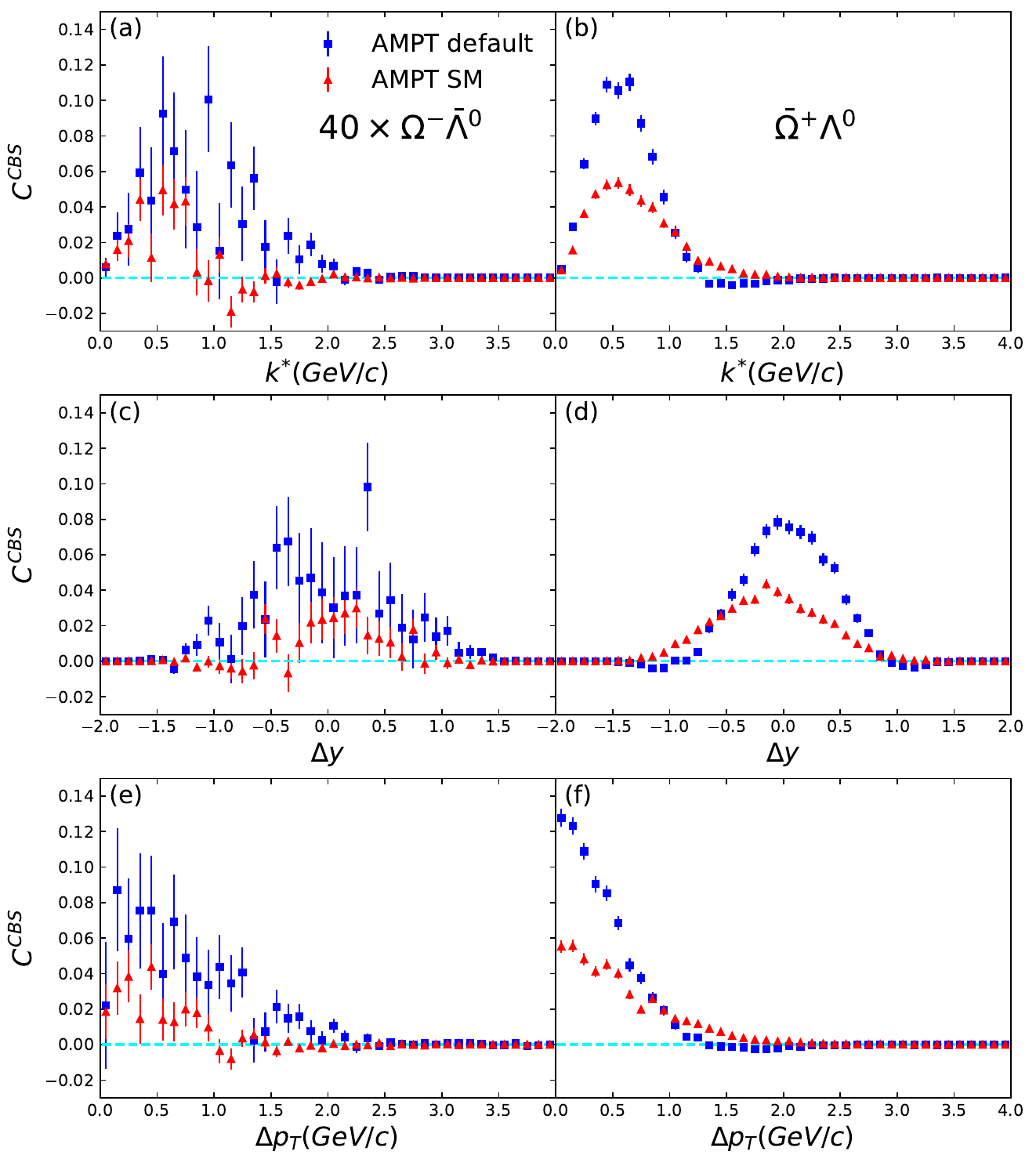}
\caption{\label{fig:8}
CBS correlations between $\Omega^-(\bar{\Omega}^{+})$ and $\bar{\Lambda}(\Lambda)$ in 0--5\%
most central Au + Au collisions at 7.7 GeV ($|\eta|<1$). The left columns show the results of $40\times C^{\rm CBS}_{\Omega^- {\bar \Lambda}}$, and the right columns, those of $C^{\rm CBS}_{\bar{\Omega}^{+}\Lambda}$.
}
\end{figure}

\begin{figure}[t]
\graphicspath{ {figures/} }
\includegraphics[width=0.48\textwidth]{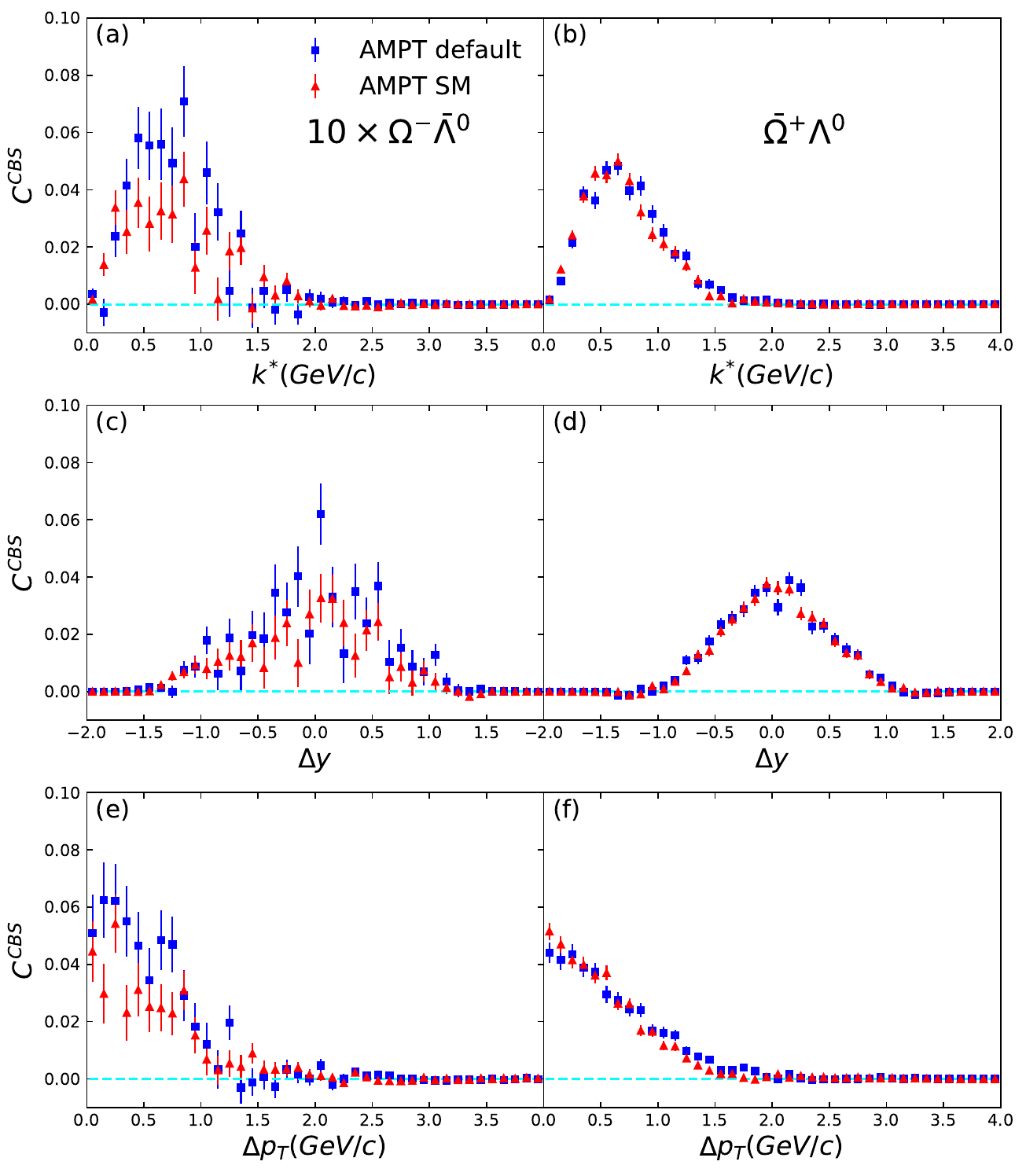}
\caption{\label{fig:9} CBS correlations between $\Omega^-(\bar{\Omega}^{+})$ and $\bar{\Lambda}(\Lambda)$ in 0--5\%
most central Au + Au collisions at 14.6 GeV ($|\eta|<1$). The left columns show the results of $10\times C^{\rm CBS}_{\Omega^- {\bar \Lambda}}$, and the right columns, those of $C^{\rm CBS}_{\bar{\Omega}^{+}\Lambda}$.
}
\end{figure}

Figure~\ref{fig:5} shows (a) $\Omega^{-}$-$K^+$, (b) $\Omega^{-}$-$\bar{\Lambda}$, and (c) $\Omega^-$-$\bar{\Xi}^+$ correlations for 0--5\% centrality Au + Au collisions at 14.6 GeV. The $\Omega^{-}$-$K^+$ correlation at 14.6 GeV seems to be less prominent than that at 7.7 GeV.
This could be explained by our normalization scheme, in which the normalization factor is the number of uncorrelated $\Omega^{-}$-$K^{+}$ pairs. More kaons are produced at higher energies, resulting in more uncorrelated kaons pairing with $\Omega$ at 14.6 GeV, diluting the correlation between the $\Omega^{-}$-$K^{+}$ due to the combinatorial background.

According to Table~\ref{table:syst_glw1}, the correlation between $\Omega$ and anti-hyperons may also be susceptible to the two $\Omega$ production scenarios. The middle and right panels of Figs.~\ref{fig:4} and \ref{fig:5} show the $\Omega^{-}$-$\bar{\Lambda}^{0}$ and $\Omega^{-}$-$\bar{\Xi}^{+}$ correlation functions using the event-mixing technique at 7.7 GeV and 14.6 GeV, respectively. At both energies, the $\Omega^{-}$-$\bar{\Lambda}^{0}$ results lack enough statistics for a definitive conclusion,  whereas some level of the $\Omega^{-}$-$\bar{\Xi}^{+}$ correlation may  exist. However, there is no significant difference between $\Omega^{-}$-$\bar{\Xi}^{+}$ and $\bar{\Omega}^{+}$-$\Xi^{-}$, similar to the $\Omega$-$K$ results. 

For both beam energies, there is no significant difference in the observed correlations between the two AMPT hadronization schemes. It seems that the correlations using the event-mixing technique are only sensitive to the kinematic region where strangeness conservation produces the $s\bar s$ pairs. The magnitude differences as shown in Table~\ref{table:syst_glw3} are, however, not quantitatively reflected in the measured correlations  due to the normalization scheme. We will explore the CBS correlations  within the same event framework and background subtraction scheme. 

Our goal is to use the $\bar{\Omega}^+$-$K^-$ correlation (Scenario 2 only) as a reference to compare with the  $\Omega^-$-$K^+$ correlation. For each of these CBS correlations, we again select events with one $\Omega^-$ or $\bar{\Omega}^+$ from 0--5\% most central collisions. As described in Eq.~(\ref{3}), the combinatorial background of the $\Omega^-$-$K^+$ correlation is modeled by the $\bar{\Omega}^+$-$K^+$ correlation based on events with one $\bar{\Omega}^+$.  $\bar{\Omega}^+$ and $K^+$ both containing $\bar s$ quarks and the narrow centrality bin make the $\bar{\Omega}^+$-$K^+$ correlation a good candidate for the combinatorial background in the $\Omega^-$-$K^+$ correlation. Similarly, we can also take the difference between $\bar{\Omega}^+$-$K^-$ and $\Omega^-$-$K^-$ to calculate the $\bar{\Omega}^+$-$K^-$ correlation, which represents contributions only from SC. To conclude, $\Omega^-$-$K^+$ includes both BNT and SC dynamics, while $\bar{\Omega}^+$-$K^-$ only includes SC. Therefore, any difference between $\Omega^-$-$K^+$ and $\bar{\Omega}^+$-$K^-$ correlations will be sensitive to the net BNT dynamics, which are absent in Scenario 2. 

Figures~\ref{fig:6} and \ref{fig:7} show the CBS correlations, for (left) $\Omega^{-}$-$K^{+}$ and (right) $\bar{\Omega}^{+}$-$K^{-}$ at 7.7 GeV and 14.6 GeV, respectively. The correlations are shown as a function of $k^*$, rapidity difference ($\Delta y$), and transverse momentum difference ($\Delta p_T$) in three rows, respectively. At 7.7 GeV, the two AMPT versions exhibit clear differences. Compared with the SM version, the default version shows a stronger $\bar{\Omega}^{+}$-$K^{-}$ correlation, presumably indicating a stronger and more localized SC. For the $\Omega^{-}$-$K^{+}$ correlation, which is sensitive to both SC and BNT dynamics, while the total magnitudes (integral) are similar between the two AMPT versions, the correlation shapes are different as a function of both $k^{*}$ and $\Delta p_{T}$. In the SM version, the coalescence formation mechanism and the BNT dynamics seem to yield a narrower correlation function between $\Omega^-$ and $K^+$.

\begin{figure}[b]
\graphicspath{ {figures/} }
\includegraphics[width=0.48\textwidth]{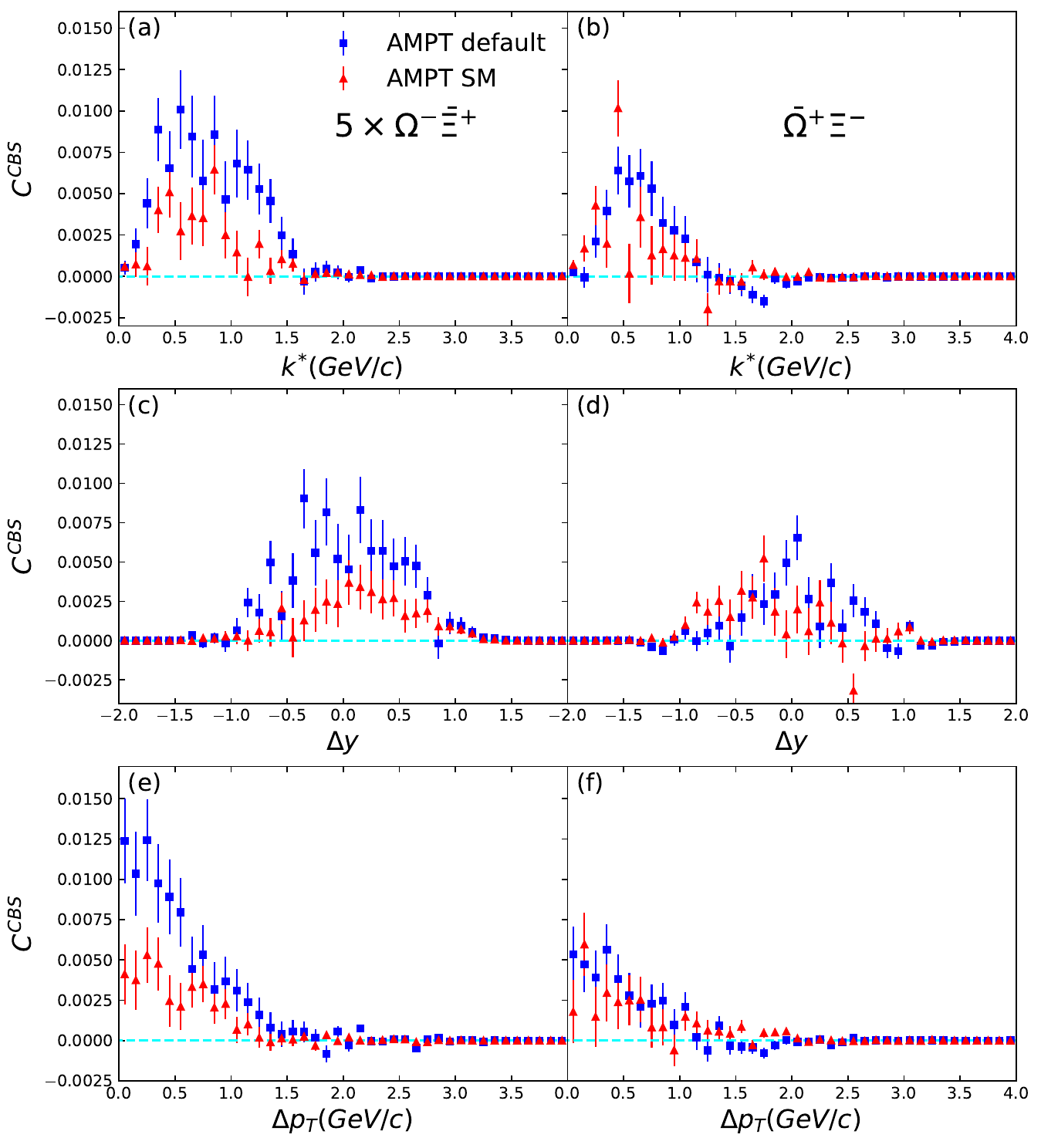}
\caption{\label{fig:10}
CBS correlations between $\Omega^-(\bar{\Omega}^{+})$ and $\bar{\Xi}^{+}(\Xi^-)$ in 0--5\%
most central Au + Au collisions at 7.7 GeV ($|\eta|<1$). The left columns show the results of $5\times C^{\rm CBS}_{\Omega^- {\bar \Xi}^+}$, and the right columns, those of $C^{\rm CBS}_{\bar{\Omega}^{+}\Xi^- }$. 
}
\end{figure}
\begin{figure}[b]
\graphicspath{ {figures/} }
\includegraphics[width=0.48\textwidth]{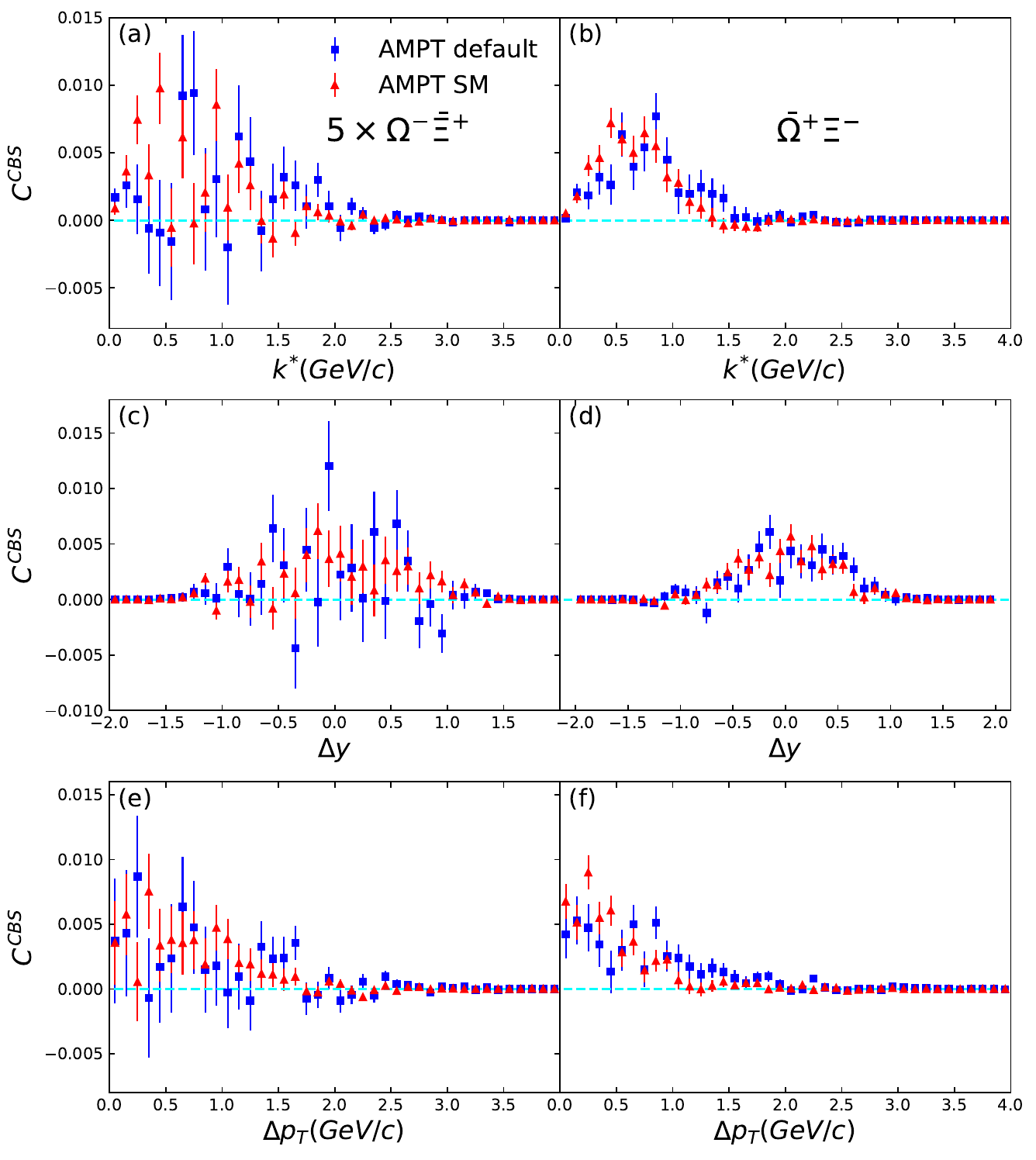}
\caption{\label{fig:11}
CBS correlations between $\Omega^-(\bar{\Omega}^{+})$ and $\bar{\Xi}^{+}(\Xi^-)$ in 0--5\%
most central Au + Au collisions at 14.6 GeV ($|\eta|<1$). The left columns show the results of $5\times C^{\rm CBS}_{\Omega^- {\bar \Xi}^+}$, and the right columns, those of $C^{\rm CBS}_{\bar{\Omega}^{+}\Xi^- }$. 
}
\end{figure}
At 14.6 GeV (Fig.~\ref{fig:7}), the difference between the two hadronization schemes is relatively small. The shape difference in the correlation as a function of $k^{*}$ and $\Delta p_{T}$ is much less prominent than at 7.7 GeV.  The CBS correlations at 7.7 GeV and 14.6 GeV suggest that the event-level $\Omega^{-}$-$K^{+}$ correlation is stronger than the $\bar{\Omega}^{+}$-$K^{-}$ one. In Scenario 2, $\Omega^{-}$ doesn't carry a net baryon number transported from colliding nuclei. Because $\bar{\Omega}^{+}$ is produced by pair production only, whereas $\Omega^{-}$ can be produced in both scenarios, the stronger $\Omega^{-}$-$K^{+}$ correlation indicates significant contributions of Scenario 1 for the $\Omega^{-}$ production at both beam energies. Particularly at 7.7 GeV, Scenario 1 seems to contribute more to the $\Omega^{-}$ production in the SM version, as the difference in correlation amplitudes between $\Omega^{-}$-$K^{+}$ and $\bar{\Omega}^{+}$-$K^{-}$ becomes noticeably larger. 

Figures~\ref{fig:8} and \ref{fig:9} show
the CBS correlations between $\Omega$ and $\bar{\Lambda}$, and their anti-particle pairs as a function of three kinematic variables in 0--5\%
most central Au + Au collisions at 7.7 GeV and 14.6 GeV, respectively. 
Besides strangeness and baryon number conservation that govern
the correlations between $\Omega^-$ and $\bar{\Lambda}$, the $\bar{\Omega}^+$-$\Lambda$ correlations are also sensitive to BNT dynamics. 
At 7.7 GeV, the default version of AMPT displays stronger correlations than the SM version.

For both AMPT versions, the $\Omega^{-}$-$\bar{\Lambda}$ correlations are much weaker than the $\bar{\Omega}^{+}$-$\Lambda$ ones, confirming that $\bar{\Omega}^{+}$ is only produced via Scenario 2. The default version also indicates a larger discrepancy between $\bar{\Omega}^{+}$-$\Lambda$ and $\Omega^{-}$-$\bar{\Lambda}$, which suggests a more prominent contribution from Scenario 2, complementary to the information on the difference between the $\Omega^{-}$-$K^{+}$ and $\bar{\Omega}^{+}$-$K^{-}$ results. At 14.6 GeV, the $\bar{\Omega}^{+}$-$\Lambda$ correlations (the right column in Fig.~\ref{fig:9}) are also higher than $\Omega^{-}$-$\bar{\Lambda}$ (the left column in Fig.~\ref{fig:9}) by about an order of magnitude. However, there seems to be no significant difference between the two AMPT versions at this energy. 



Figures~\ref{fig:10} and \ref{fig:11} show the CBS correlations between  $\Omega$ and $\bar{\Xi}$ as a function of three kinematic variables in 0--5\%
most central Au + Au collisions at 7.7 GeV and 14.6 GeV, respectively. At both beam energies, the stronger $\bar{\Omega}^{+}$-$\Xi^{-}$ correlations relative to $\Omega^{-}$-$\bar{\Xi}^{+}$ also imply that Scenario 2 provides a larger contribution to the $\bar{\Omega}^{+}$ production than to $\Omega^{-}$. At 7.7 GeV, the difference in the $\Omega^{-}$-$\bar{\Xi}^{+}$ correlations between the two AMPT versions suggests that the default version generates a larger contribution from Scenario 2 to the $\Omega^{-}$ production. At 14.6 GeV, the $\bar{\Omega}^{+}$-$\Xi^{-}$ correlations are still stronger than $\Omega^{-}$-$\bar{\Xi}^{+}$, but no significant difference appears between the two versions.


\section{Summary}

The $\Omega$ production in nuclear collisions at the RHIC BES involves dynamics of baryon number transport, strangeness conservation, and baryon number conservation. To investigate these effects, we have used the AMPT model with both the default and the string melting versions to simulate central Au+Au  collisions at 7.7 and 14.6 GeV, and showed that $\Omega^-$-$K^+$ and $\Omega^-$-anti-hyperon correlations
are sensitive to the dynamics. In particular, we have considered two generic $\Omega$ production scenarios, one with three $\bar s$ quarks in kaons (Scenario 1) and $\Omega$ carrying a net baryon number, and the other baryon pair production of $\Omega$ and anti-baryon with $\bar s$ quarks in anti-hyperon and kaons (Scenario 2) . Both scenarios are constrained by strangeness and baryon number conservation, and only the former is sensitive to net baryon number transport. The AMPT simulations show that in  Au+Au collisions, both scenarios contribute to the $\Omega$ production, and Scenario 1 becomes  more important from 14.6 to 7.7 GeV. The shape of the correlations can also be sensitive to the hadronization schemes in the default and the string melting versions of the AMPT model. Experimental measurements of these correlations and comparisons with our AMPT simulation results could greatly advance our understanding of baryon transport dynamics and effects of strangeness and baryon number conservation on the $\Omega$ production, and possibly enable future experimental studies for exotic baryon transport mechanisms such as gluon junction interactions.




%

\end{document}